# Layer-mediated tuning of spin and valley physics in stacked tetragonal altermagnetic bilayers

Jianke Tian, Xiaowen Zhou, Gui-Bin Liu*

*Centre for Quantum Physics, Key Laboratory of Advanced Optoelectronic Quantum Architecture and Measurement (MOE),*
*School of Physics, Beijing Institute of Technology, Beijing 100081, China, Beijing Key Lab of Nanophotonics and*
*Ultrafine Optoelectronic Systems, School of Physics, Beijing Institute of Technology, Beijing 100081, China, and*
*Beijing Institute of Technology, Zhuhai 519000, China*

As an emerging magnetic phase, altermagnets (AMs) with collinear compensated magnetism in real space and alternating spin splitting in the band structure have attracted widespread attention. Here, based on first-principles calculations, we demonstrate that the layer stacking imposes symmetry constraints on the spin and valley degrees of freedom (DOFs) in an AM bilayer composed of two tetragonal altermagnetic monolayers, thereby enabling the tuning of these DOFs through interlayer sliding as well as by an external electric field. Using several representative AM bilayers, we reveal that the $[C_2\|\mathcal{P}]$ and $[C_2\|\mathcal{M}_z]$ symmetries intrinsically enforce spin degeneracy, while the coupling between spin and layer DOFs establishes a general framework for achieving electric field control of spin states. Appropriate interlayer sliding breaks the $[C_2\|\mathcal{M}_d]$ symmetry of AM bilayers, thereby giving rise to a spontaneous valley splitting and driving a transition to a fully compensated ferrimagnetic state. Furthermore, owing to the tunable valley splitting induced by interlayer sliding, enhanced tunneling magnetoresistance (TMR) can be realized by AM bilayers. This work highlights the intrinsic correlation among spin, valley, and layer DOFs, offering symmetry-based design principles for layer-based spintronic and valleytronic devices.

*Corresponding author. *E-mail address*: gbliu@bit.edu.cn (G.-B. Liu).

## I. INTRODUCTION

Altermagnetism has emerged as a novel third class of collinear magnetism, characterized by opposite-spin sublattices connected through real-space-rotation symmetry rather than spatial inversion ($\mathcal{P}$) and/or translation operation [1–11]. The unconventional nature of altermagnetic class is that the rotation symmetry connecting the opposite-spin sublattices endows altermagnets (AMs) with antiferromagnetic-like compensated magnetic order and vanishing net magnetization in real space, while simultaneously exhibiting ferromagnetic-like eV-scale nonrelativistic spin splittings in momentum space [12–19]. This unique combination of antiferromagnets and ferromagnets properties makes AM to show ultrafast switching dynamics and highly spin-polarization currents, greatly enriching fields ranging from spintronics to piezotronics [20–25]. Although several AM, including CrSb [26], $Mn_5Si_3$ [27] and MnTe [28], have been realized theoretically and experimentally, reports on the candidates for two-dimensional (2D) AM and on the manipulation of charge, spin and valley degrees of freedom (DOFs) in 2D AM remain scarce [29–31].

2D van der Waals (vdW) materials with an extra layer DOF have attracted significant attention due to their weak interlayer vdW interactions [32–41]. Monolayer materials sensitive to external stimuli can be readily stacked to form vdW systems, exhibiting rich physical properties. The twisted antiferromagnets bilayer have been predicted to be promising candidates for realizing AM in 2D systems [7,42–45]. However, even in the presence of spin splitting, the intervalley energy degeneracy in AM limits access to valley-resolved transport responses, such as the anomalous valley Hall effect (AVHE) and valley-polarized spin currents. Recently, studies on valley splitting states have expanded from ferromagnets systems with spontaneous valley splitting to AM, including the $V_2Se_2O$ [5], CrX (X = O, S) [36,46], $Fe_2WZ_4$ (Z = S, Se) [47], twisted VOBr [43], twisted $NiZrI_6$ [45] and twisted $NiCl_2$ [48]. Strain engineering has emerged as the primary route for inducing valley splitting in these AM monolayers and twisted vdW stackings [39,49,50]. A natural question arises: in addition to strain engineering, are there other methods such as sliding engineering available for inducing valley splitting? Likewise, could the interlayer magnetic configuration provide a means to control spin behavior within the system? Exploring these possibilities calls for a thorough understanding of the potential coupling among spin, valley, and layer DOFs in vdW systems, which is essential for both theoretical and experimental studies in condensed matter physics.

In this work, based on symmetry analysis and first-principles calculations, we reveal a layer-mediated mechanism for controlling spin and valley DOFs in AM bilayers via interlayer sliding, magnetic reconfiguration or an external electric field ($E_z$). Note that the term "AM bilayer" used in this paper refers merely to a bilayer constructed from two AM monolayers, regardless of whether the bilayer as a whole preserves altermagnetism. In AM bilayers, spin and valley states are constrained by symmetry, and breaking the symmetry naturally becomes an effective tuning method. We concentrate mainly on two kinds of symmetry constraints: (i) spin degeneracy is protected by either $[C_2\|\mathcal{P}]$ or $[C_2\|\mathcal{M}_z]$ symmetry, and it can be lifted by magnetic reconfiguration or an external electric field; (ii) valley degeneracy is protected by $[C_2\|\mathcal{M}_d]$ symmetry, and interlayer sliding can break this symmetry to induce valley splitting. Moreover, $[C_2\|\mathcal{M}_d]$ and $[C_2\|\mathcal{P}]/[C_2\|\mathcal{M}_z]$ impose independent symmetry constraints, and hence combining them enables highly tunable spin splitting and valley splitting. Our theoretical investigation suggests that symmetry-based control may provide a promising avenue for combining layer, spin, and valley functionalities.

## II. COMPUTATIONAL DETAILS

All structural optimization and electronic structure calculations are performed on the basis of density functional theory (DFT) by employing the Vienna *ab initio* simulation package (VASP) [51,52]. The projector augmented wave (PAW) method [53] is used to describe the interaction between electrons and ions. A generalized gradient approximation (GGA) of the Perdew-Burke-Ernzerhof (PBE) form is used as exchange-correlation functional [54]. The kinetic energy cutoff is set to be 500 eV and a 20 Å vacuum layer in the $z$ direction is used to eliminate interactions between adjacent layers. The 11×11×1 Γ-centered k meshes of 2D Brillouin zone (BZ) are used. The total energy convergence criterion is set to be $10^{-6}$ eV, and force convergence criteria of less than 0.01 eV/Å on each atom are used to attain reliable results. To properly account for the localized V-3$d$, Fe-3$d$ orbitals, the effective Hubbard values $U$ were set to 4.3 eV for $V_2S_2O$, $V_2Se_2O$, $V_2SSeO$ [5], and to 2 eV for $Fe_2MoS_4$ [35]. The phonon dispersion is based on a 4×4×1 supercell by using the PHONOPY code [55]. The *ab initio* molecular dynamic (AIMD) simulations adopt the *NVT* ensemble based on the Nosé-Hoover thermostat [56]. To describe the interlayer interaction of bilayer, the DFT-D3 method is employed to deal with vdW interaction [57]. The crystal structures were visualized using VESTA software [58]. The VASP data are processed by VASPKIT code [59].

## III. RESULTS AND DISCUSSIONS

### A. Symmetry analysis for AM bilayer

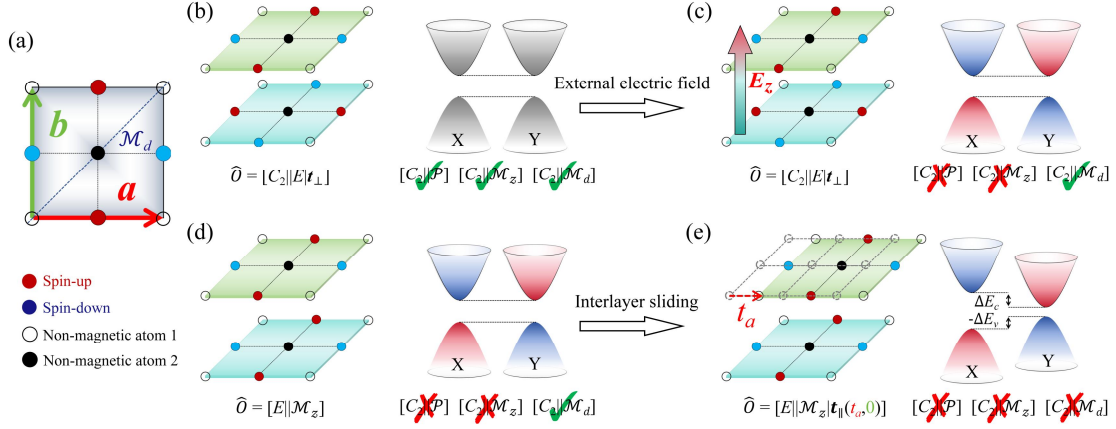

FIG. 1. (a) Schematic illustration of the cell of a tetragonal AM monolayer. (b) The AM bilayer constructed via the stacking operation $\hat{O}=[C_2\|E|t_\perp]$, and its corresponding spin-degenerate band structure protected independently by the $[C_2\|\mathcal{P}]$ and $[C_2\|\mathcal{M}_z]$ symmetries. (c) Bilayer with electric field and its spin-splitting band structure. (d) The AM bilayer stacked via $\hat{O}=[E\|\mathcal{M}_z]$, and its corresponding spin-splitting band structure. (e) Bilayer with interlayer sliding and its valley-splitting band structure. "✓" ("✗") denotes the presence (absence) of a certain symmetry in the AM bilayer.

To comprehend the layer-mediated spin and valley tuning in AM bilayer, it is essential to use spin space group (SSG) for symmetry analysis. The emerging altermagnetic phase belongs to nonrelativistic spin-collinear systems, and the nontrivial collinear SSG that describes AM is given by [1,2]:

$$\mathbf{S}=[E\|\mathbf{H}]+[C_2\|\mathbf{G}\text{-}\mathbf{H}]. \tag{1}$$

The set **H** denotes a halving subgroup of crystallographic space group **G**, and **G-H** denotes the coset of **H**. A general SSG operation is denoted as $[S\|R|t]$, where the symbol $\|$ emphasizes that the spin rotation $S$ and the spatial operation $\{R|t\}$ act independently. For short, we denote $[S\|R|t]$ as $[S\|R]$ when $t=0$. In Eq. (1), $S=C_2$ represents a 180° spin-rotation around an axis perpendicular to the spin. The symmetries in **H** guarantee that each sublattice with a given spin is preserved, whereas those in the coset **G-H** interchange the spin-up and spin-down sublattices. Generally, if a collinear system has $[S\|R|t]$ symmetry, its energy dispersion $E_\sigma(\mathbf{k})$ is constrained by this symmetry as follow

$$[S\|R|t]E_\sigma(\mathbf{k})=E_{S\sigma}(R\mathbf{k})=E_\sigma(\mathbf{k}), \tag{2}$$

in which $\sigma$ is the spin index ↑ or ↓ and can be flipped by the $C_2$ or time reversal $\mathcal{T}$. Note that $\mathcal{T}$ acts globally on both the spin and spatial DOFs and hence can be written as $[\mathcal{T}\|\mathcal{T}]$ in the SSG notation. As is known, collinear magnets always possess the

$[C_2\mathcal{T}\|\mathcal{T}]$ symmetry [1,2,4]. Accordingly, if the system additionally preserves the $[C_2\|\mathcal{P}]$ symmetry, spin degeneracy $E_\uparrow(\bm{k})=E_\downarrow(\bm{k})$ occurs at each $\bm{k}$, which is enforced by $[C_2\|\mathcal{P}][C_2\mathcal{T}\|\mathcal{T}]=[\mathcal{T}\|\mathcal{P}\mathcal{T}]$. In addition, for a layered material, the presence of the $[C_2\|\mathcal{M}_z]$ symmetry alone guarantees spin degeneracy across the entire 2D BZ [45,48]. In a word, in order to obtain spin splitting, one has to break both $[C_2\|\mathcal{P}]$ and $[C_2\|\mathcal{M}_z]$ symmetries in a spin-collinear layered system.

Each bilayer studied in this work is constructed by stacking two AM monolayers through a stacking operation $\hat{O}$. Let $L_b$ denote the bottom layer. The top layer is thus generated by applying $\hat{O}$ to $L_b$, such that $L_b+\hat{O}L_b$ is the resulting bilayer. Accordingly, even when constructed from the same $L_b$, different operations $\hat{O}$ lead to distinct bilayers with different symmetries. Hereafter, we label each bilayer structure by the corresponding operation $\hat{O}$ indicated in parentheses. Note that $\hat{O}$ merely labels the construction of the bilayer and does not imply that the bilayer itself is symmetric under $\hat{O}$. The monolayers we used are all tetragonal AMs with Néel-type magnetic order. They all have $[C_2\|\mathcal{M}_d]$ symmetry which relates the spin-up and spin-down sublattices, where $\mathcal{M}_d$ denotes the vertical diagonal mirror, with the mirror plane normal to the $[1\bar{1}0]$ direction, as shown in Fig. 1(a). As for the stacking operation, $\hat{O}=[C_2\|E|\bm{t}_\perp]$, $\hat{O}=[E\|E|\bm{t}_\perp]$, $\hat{O}=[C_2\|\mathcal{M}_z|\bm{t}_\|]$, $\hat{O}=[E\|\mathcal{M}_z|\bm{t}_\|]$ are mainly involved in this work, where $\bm{t}_\perp$ and $\bm{t}_\|$ are translation vectors perpendicular and parallel to the layer, respectively. $|\bm{t}_\perp|$ is the interlayer distance and $\bm{t}_\|=t_a\bm{a}+t_b\bm{b}$ is also denoted as $\bm{t}_\|(t_a,t_b)$ [see Fig. S1 of Supplemental Material (SM)].

AM bilayers constructed via $\hat{O}=[C_2\|E|\bm{t}_\perp]$ and $\hat{O}=[E\|\mathcal{M}_z]$ are demonstrated in Fig. 1(b) and Fig. 1(d), respectively. In Fig. 1(b), we assume that the monolayer has $[E\|\mathcal{M}_z]$ and $[E\|\mathcal{P}]$ symmetries (such as $V_2S_2O$), and hence the resulting bilayer has both $[C_2\|\mathcal{M}_z]$ and $[C_2\|\mathcal{P}]$ symmetries. Accordingly, the energy bands of the bilayer are spin degenerate; that is, this AM bilayer does not, in fact, preserve altermagnetism. To restore altermagnetism, one can apply external electric field to break both $[C_2\|\mathcal{M}_z]$ and $[C_2\|\mathcal{P}]$ symmetries, as shown in Fig. 1(c). In Fig. 1(d), the bilayer constructed via $\hat{O}=[E\|\mathcal{M}_z]$ intrinsically lacks the $[C_2\|\mathcal{M}_z]$ and $[C_2\|\mathcal{P}]$ symmetries. As a result, its energy bands exhibit spin splitting, indicating that altermagnetism is preserved.

In a periodic crystal, the wave vector star $\mathcal{S}(\bm{k})$ of a given wave vector $\bm{k}$ is defined as the set of all symmetry-related wave vectors generated under the action of the symmetry operations Q

$$\mathcal{S}(k) = \{Qk \mid Q \in \mathbf{G_0}\} \qquad (3)$$

where $\mathbf{G_0}$ is the spin point group of SSG. According to Eq. (2), each pair of wave vectors in the same star $\mathcal{S}(k)$ are symmetry-related and result in energy degeneracy between the points in the BZ, i.e., $E(k_i)=E(k_j)$, $\forall\, k_i, k_j \in \mathcal{S}(k)$ (spin index omitted). The bilayers in Figs. 1(b)-1(d) possess the $[C_2\|\mathcal{M}_d]$ symmetry, which relates the wave vectors $k_X$ and $k_Y$ of the X and Y valleys. $k_X$ and $k_Y$ belong to the same star, therefore valley degeneracy $E_{v/c}(k_X)=E_{v/c}(k_Y)$ exists, where the subscripts $v$ and $c$ represent the valence band maximum (VBM) and conduction band minimum (CBM), respectively. In bilayers, the intrinsic layer DOF provides a layer-mediated route for reassigning previously symmetry-related wave vectors into different wave vector stars, resulting in valley splitting $E_{v/c}(k_X) \neq E_{v/c}(k_Y)$. The magnitude of the valley splitting is defined as $\Delta E_{v/c}=E_{v/c}(k_X)-E_{v/c}(k_Y)$.

Previous studies have employed strain engineering as an effective methods to break the $[C_2\|\mathcal{M}_d]$ symmetry and consequently achieved the valley splitting in AM systems [39,60,61]. Inspired by the above symmetry analysis, we propose that interlayer sliding engineering provides an alternative route beyond strain engineering to manipulate the valley DOF in AM bilayer. Starting from the AM bilayer in Fig. 1(d), we keep the bottom layer fixed and translate the top layer by $t_a$ along the $\boldsymbol{a}$ direction, thereby generating the interlayer-sliding AM bilayer ($\widehat{O}=[E\|\mathcal{M}_z|t_\|(t_a,0)]$) shown in Fig. 1(e). The interlayer sliding breaks the $[C_2\|\mathcal{M}_d]$ symmetry, thereby inducing a valley splitting. In fact, both spin splitting and valley splitting exist in this interlayer-sliding bilayer. If interlayer sliding is applied to the spin-degenerate bilayer in Fig. 1(b), valley splitting emerges while spin degeneracy remains. This is because, although the interlayer sliding breaks the $[C_2\|\mathcal{M}_z]$ symmetry, the presence of $[C_2\|\mathcal{P}]$ still enforces spin-degenerate bands [see Fig. S2 of SM]. In summary, the introduction of the layer DOF gives rise to a layer-mediated mechanism for tuning spin and valley DOFs, wherein the $[C_2\|\mathcal{P}]/[C_2\|\mathcal{M}_z]$ and $[C_2\|\mathcal{M}_d]$ symmetries independently govern the spin and valley behavior, respectively.

## B. The layer-mediated spin tuning in AM bilayer

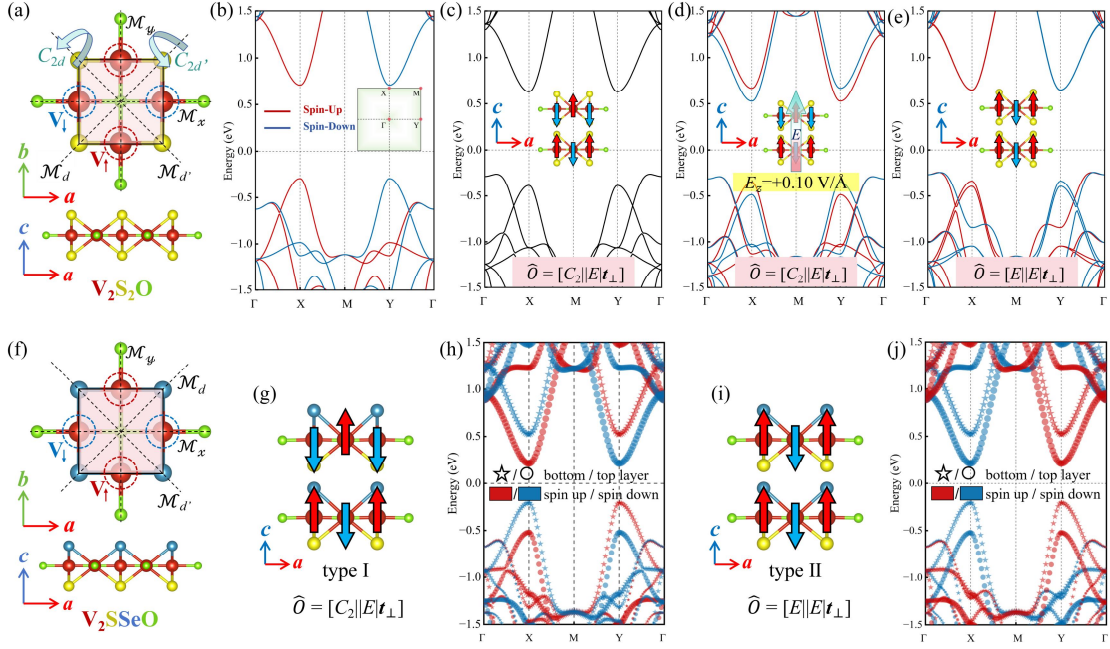

FIG. 2. (a) Crystal structure of $V_2S_2O$ monolayer. (b) Band structure of $V_2S_2O$ monolayer. (c) Spin-degenerate band structure of $V_2S_2O$ bilayer ($\hat{O}=[C_2\|E|t_\perp]$) protected by the $[C_2\|\mathcal{P}]/[C_2\|\mathcal{M}_z]$ symmetry. The magnetization of the top and bottom layers is oriented antiparallel. (d) Spin-splitting band structure of $V_2S_2O$ bilayer ($\hat{O}=[C_2\|E|t_\perp]$) under an external electric field $E_z=0.10$ V/Å, where the $[C_2\|\mathcal{P}]/[C_2\|\mathcal{M}_z]$ symmetry is broken. (e) Spin-splitting band structure of $V_2S_2O$ bilayer ($\hat{O}=[E\|E|t_\perp]$) without $[C_2\|\mathcal{P}]/[C_2\|\mathcal{M}_z]$ symmetry. The magnetization of the top and bottom layers is aligned parallel. (f) Crystal structure of Janus $V_2SSeO$ monolayer. (g) Stacking configuration and (h) spin-layer-resolved band structure of type-I $V_2SSeO$ bilayer ($\hat{O}=[C_2\|E|t_\perp]$). (i) Stacking configuration and (j) spin-layer-resolved band structure of type-II $V_2SSeO$ bilayer ($\hat{O}=[E\|E|t_\perp]$). Red and blue arrows denote up and down spins, respectively.

Here, we examplify the symmetry analysis of the spin behavior in 2D tetragonal AM bilayers using DFT simulations. Firstly, the possible five magnetic configurations [see Fig. S3 of SM], i.e. non-magnetism (NM), ferromagnetism (FM), Néel-type, Zigzag-type and Stripy-type magnetic orders, are considered to evaluate the magnetic ground state of $V_2S_2O$, $V_2Se_2O$, $V_2SSeO$ and $Fe_2MoS_4$ monolayers. Calculated results indicate that all monolayers exhibit the Néel-type magnetic ground state, with optimized lattice constants listed in Table S1. The dynamical and thermal stabilities of these structures have been verified through the phonon spectrum and AIMD calculation [refer to Fig. S4 of SM], and the mechanical stability is confirmed by the corresponding elastic constants listed in Table. S1 of SM, which satisfy the Born-Huang criteria ($C_{11}>0$, $C_{22}>0$, $C_{66}>0$ and $C_{11}-C_{12}>0$) [62].

We use the $V_2S_2O$ and $V_2SSeO$ systems as the representative examples to discuss the above proposal in detail [for more details on other AM monolayers, see Sec. IV of

SM]. Fig. 2(a) illustrates the V$_2$S$_2$O monolayer with SSG $P4^m/m^1m^1m^m$ (No. 123.1.2.7.L) [63]. The superscripts "1" and "m" in the SSG notations denote the identity and spin-reversing operations, respectively. Based on Eq. (1), the sets **H** and **G-H** of V$_2$S$_2$O monolayer contain the symmetry elements {$E$, $\mathcal{P}$, $C_{2x}$, $C_{2y}$, $C_{2z}$, $\mathcal{M}_x$, $\mathcal{M}_y$, $\mathcal{M}_z$} and {$C_{4z}^+$, $C_{4z}^-$, $S_{4z}^+$, $S_{4z}^-$, $\mathcal{M}_d$, $\mathcal{M}_{d'}$, $C_{2d}$, $C_{2d'}$}, respectively. As shown in Fig. 2(b), the X and Y valleys host opposite spin channels while remaining energetically degenerate, consistent with the altermagnetic character of V$_2$S$_2$O monolayer. We first focus on the V$_2$S$_2$O bilayer constructed via $\widehat{O}=[C_2\|E|t_\perp]$, as show in Fig. 1(b). This V$_2$S$_2$O bilayer retains spin degeneracy, as shown in Fig. 2(c), which is enforced by the $[C_2\|\mathcal{P}]$ and $[C_2\|\mathcal{M}_z]$ symmetries. It is well known that an external electric field directly couples to the layer DOF, generating a layer-dependent electrostatic potential that breaks the $\mathcal{P}$ and $\mathcal{M}_z$ spatial symmetries in vdW bilayer. As a result, an external electric field $E_z$ (0.10 V/Å) breaks the $[C_2\|\mathcal{P}]$ and $[C_2\|\mathcal{M}_z]$ symmetries of the V$_2$S$_2$O bilayer, leading to spin splitting of the bands, as shown in Fig. 2(d). Experimentally, the manipulation of the Néel vector can be achieved through spin-orbit torques, which motivates us to explore spin splitting by modifying the magnetic configuration of the system [64,65]. By changing the stacking operation from $\widehat{O}=[C_2\|E|t_\perp]$ to $\widehat{O}=[E\|E|t_\perp]$, the resulting bilayer in Fig. 2(e) exhibits a different magnetic configuration compared to that in Fig. 2(c). The parallel spin alignment between the bottom and top layers of the bilayer ($\widehat{O}=[E\|E|t_\perp]$) breaks both the $[C_2\|\mathcal{P}]$ and $[C_2\|\mathcal{M}_z]$ symmetries, even in the absence of an electric field, thereby also inducing a spin-splitting band structure.

Then, we further verified the crucial role of the layer DOF in effectively tuning the spin splitting in AM bilayer. As shown in Fig. 2(f), the Janus V$_2$SSeO monolayer intrinsically breaks the $\mathcal{P}$ and $\mathcal{M}_z$ spatial symmetries. Therefore, the V$_2$SSeO bilayers of both the type-I ($\widehat{O}=[C_2\|E|t_\perp]$) and type-II ($\widehat{O}=[E\|E|t_\perp]$) magnetic configurations [see Figs. 2(g) and 2(i)] exhibit spin-splitting band structures [see Figs. 2(h) and 2(j)]. By comparing the type-I and type-II magnetic configurations, one can see that flipping the spins of the magnetic atoms in the top layer [cf. Figs. 2(g) and 2(i)] leads to a corresponding interchange of their contributions to the spin-up and spin-down channels in the top-layer-projected bands [see Figs. 2(h) and 2(j)]. Alternatively, in addition to changing the magnetic configuration, an external electric field $E_z$ can also reverse the spin channels, as shown in Fig. 3(a). Electric fields applied in the $+z$ and $-z$ directions induce opposite spin splitting at both the VBM and CBM. Here, the magnitude of spin splitting of VBM/CBM is defined as $\Delta E_{\mathrm{spin}}=E_{\uparrow X(v/c)}-E_{\downarrow X(v/c)}$. As shown in Fig. 3(b), by applying an accessible electric field $E_z$ ranging from -0.15 to 0.15 V/Å, we reveal its capability to modulate the spin splitting in AM bilayer. In the absence of electric field, the AM bilayer ($\widehat{O}=[C_2\|\mathcal{M}_z|\boldsymbol{t}_\|(1/2,1/2)]$) remains spin degenerate due to the protection of the $[C_2\|\mathcal{P}]$ symmetry. In contrast, electric fields of opposite directions generate opposite spin splittings, and an accessible electric field of 0.10 V/Å can induce a spin splitting $\Delta E_{\mathrm{spin}}$ exceeding 300 meV at the VBM of $Fe_2MoS_4$ bilayer (more details are provided in Sec. V of SM).

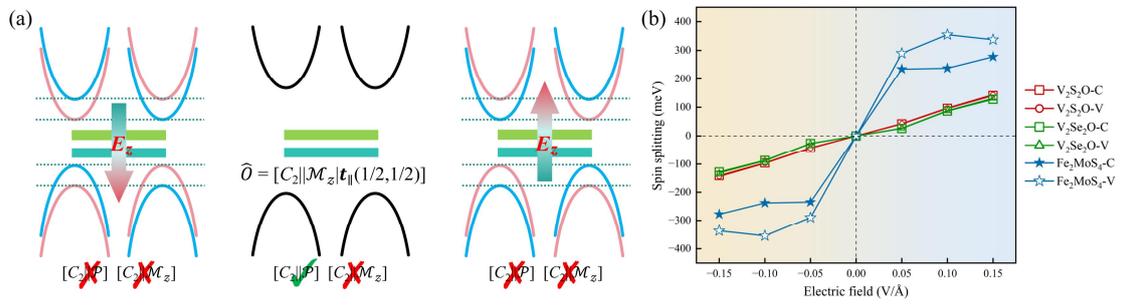

FIG. 3. (a) Schematic diagram of spin reversal by electric field. Red and blue lines correspond to spin-up and spin-down bands, respectively, and black lines indicate spin-degenerate bands. (b) Spin splitting in the VBM and CBM of $Fe_2MoS_4$, $V_2S_2O$, $V_2Se_2O$ bilayers ($\widehat{O}=[C_2\|\mathcal{M}_z|\boldsymbol{t}_\|(1/2,1/2)]$), modulated by external electric field.

## C. The layer-mediated valley tuning in AM bilayer

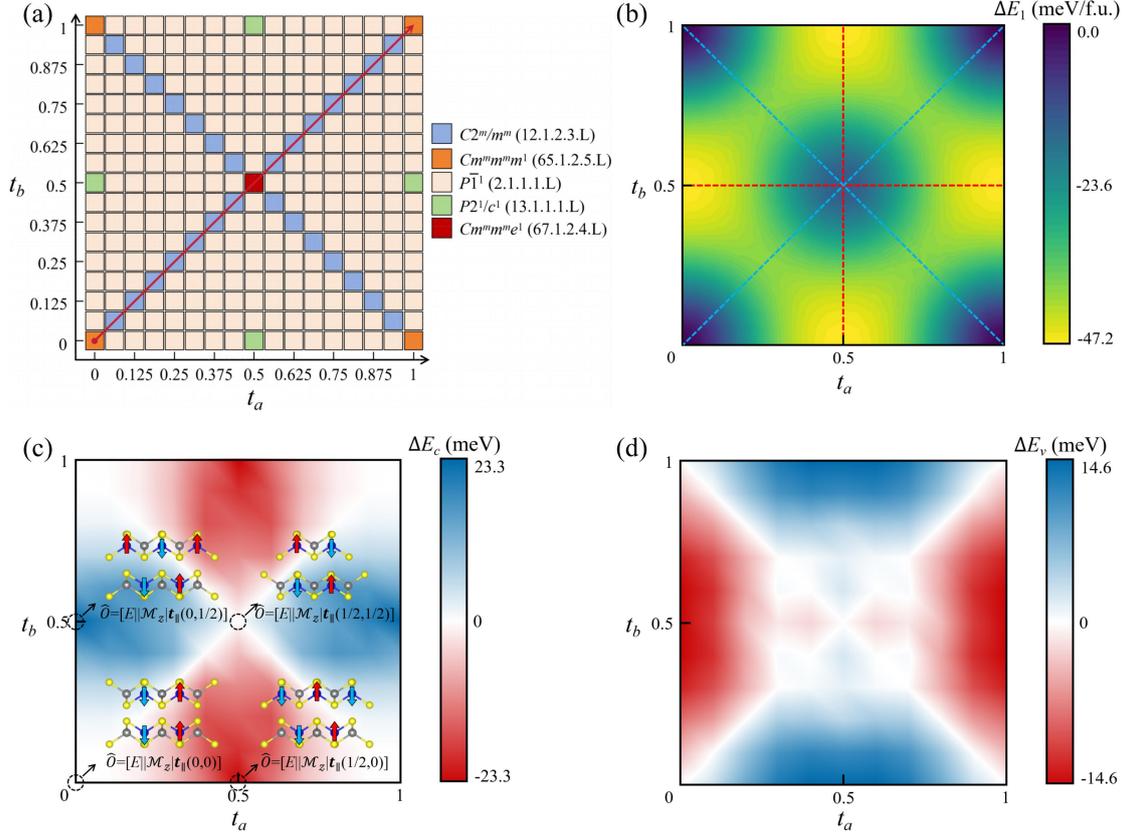

FIG. 4. Sliding-dependent properties of the Fe$_2$MoS$_4$ bilayer ($\hat{O}=[E\|\mathcal{M}_z|t_\|(t_a,t_b)]$). (a) The SSGs. The superscripts "1" and "$m$" in the SSG notations denote the identity and spin-reversing operations, respectively. (b) The energy difference relative to the energy of the Fe$_2$MoS$_4$ bilayer without interlayer sliding. Valley splitting at (c) CBM and (d) VBM. The insets of (c) show the representative bilayer configurations at high-symmetry sliding positions.

After identifying $[C_2\|\mathcal{P}]$ and $[C_2\|\mathcal{M}_z]$ symmetries as the key factor tuning the spin splitting, we proceed to explore the layer-mediated valley tuning mechanism in AM bilayers. As a concrete example, we study the Fe$_2$MoS$_4$ bilayer constructed via $\hat{O}=[E\|\mathcal{M}_z|t_\|(t_a,t_b)]$, which lacks $[C_2\|\mathcal{P}]$ and $[C_2\|\mathcal{M}_z]$ symmetries. Fe$_2$MoS$_4$ is a representative member of the widely investigated M$_2$XY$_4$ family (M = Fe, Mn, Co, X = Mo, W, Y = S, Se, Te) [47,61,65]. Based on this system, we elucidate the symmetry constrains imposed by $[C_2\|\mathcal{M}_d]$ on the valley splitting. Interlayer sliding modifies the stacking registry of the AM bilayer, thereby inducing a corresponding evolution of the SSG. The SSGs of all the Fe$_2$MoS$_4$ bilayers with sliding are given in Fig. 4(a), while the configurations of the representative Fe$_2$MoS$_4$ bilayers at high-symmetry sliding positions are illustrated in the insets of Fig. 4(c). Five distinct SSGs can be identified, and the Fe$_2$MoS$_4$ bilayers with the SSG $P2^1/c^1$ possess the lowest energy, corresponding to the sliding vectors $t_\| = \frac{1}{2}a$ or $\frac{1}{2}b$ [see Fig. 4(b)].

The SSG of the Fe$_2$MoS$_4$ bilayer without interlayer sliding is $Cm^mm^mm^1$, and the sets **H** and **G-H** contain the symmetry elements {$E$, $\mathcal{P}$, $C_{2z}$, $\mathcal{M}_z$} and {$C_{2d}$, $C_{2d'}$, $\mathcal{M}_d$, $\mathcal{M}_{d'}$}, respectively. Under the interlayer sliding with $t_\parallel(t_a,t_b)$: (i) when $t_a = t_b$, the SSG of Fe$_2$MoS$_4$ bilayer evolves from $Cm^mm^mm^1$ to $C2^m/m^m$ [**H**={$E$, $[\mathcal{P}|t_\parallel(t_a,t_a)]$}, **G-H**={$[C_{2d'}|t_\parallel(t_a,t_a)]$, $\mathcal{M}_d$}], and subsequently to $Cm^mm^me^1$ [**H**={$E$, $[\mathcal{P}|t_\parallel(1/2,1/2)]$, $C_{2z}$, $[\mathcal{M}_z|t_\parallel(1/2,1/2)]$}, **G-H**={$[C_{2d}|t_\parallel(1/2,1/2)]$, $[C_{2d'}|t_\parallel(1/2,1/2)]$, $\mathcal{M}_d$, $\mathcal{M}_{d'}$}]; (ii) when $t_a \neq t_b$, the SSG of Fe$_2$MoS$_4$ bilayer evolves from $Cm^mm^mm^1$ to $P\bar{1}^1$ [**H**={$E$, $[\mathcal{P}|t_\parallel(t_a,t_b)]$}, **G-H**=∅] and further to $P2^1/c^1$ [**H**={$E$, $[\mathcal{P}|\tau]$, $C_{2z}$, $[\mathcal{M}_z|\tau]$} where $\tau = \frac{1}{2}a$ or $\frac{1}{2}b$, **G-H**=∅]. As shown in Figs. 4 and 5, the absence of $\mathcal{M}_d$ in the **G-H** set, reflecting the lack of the $[C_2\|\mathcal{M}_d]$ symmetry in the Fe$_2$MoS$_4$ bilayer, gives rise to the layer-mediated valley splitting tuning by interlayer sliding [see Sec. VI and VII of SM for more details on other bilayers].

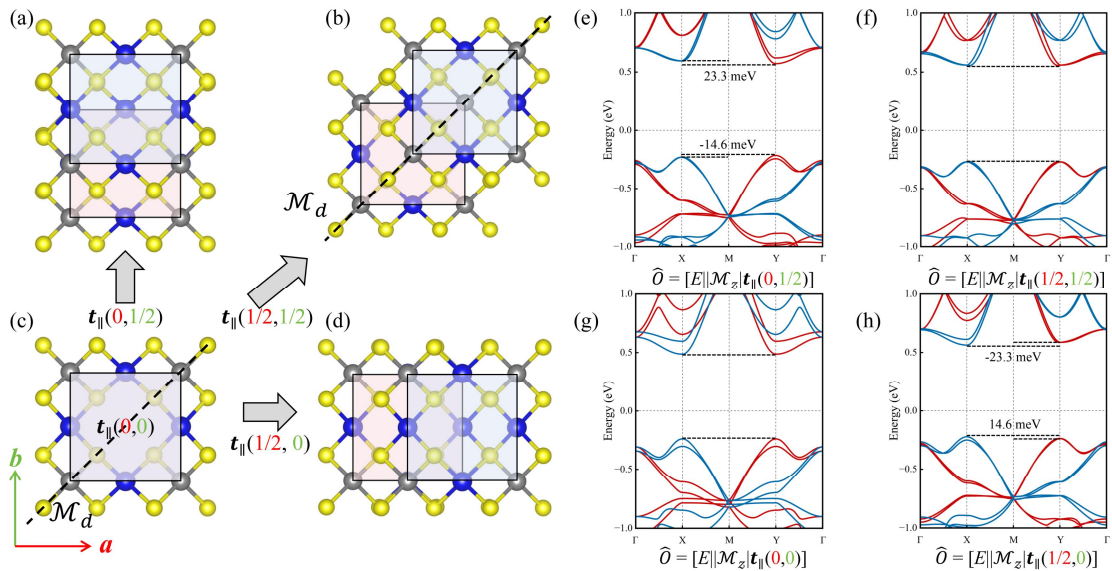

FIG. 5. Schematic crystal and band structures of Fe$_2$MoS$_4$ bilayers ($\hat{O}=[E\|\mathcal{M}_z|t_\parallel(t_a,t_b)]$) under different interlayer sliding. (a)-(d) Crystal structures for $t_\parallel$ = (0,1/2), (1/2,1/2), (0,0) and (1/2,0), respectively, while (e)-(h) display the corresponding spin-splitting and valley-splitting band structures.

In contrast to the diagonal-sliding Fe$_2$MoS$_4$ bilayer ($t_a=t_b$) whose SSG satisfies Eq. (1), the **G-H** set is empty for the non-diagonal-sliding Fe$_2$MoS$_4$ bilayer ($t_a \neq t_b$). This implies that, in a strict sense, the non-diagonal-sliding Fe$_2$MoS$_4$ bilayer exhibiting valley splitting cannot be classified as an AM, since no symmetry operation exists to exchange the spin-up and spin-down sublattices. Accordingly, the diagonal-sliding Fe$_2$MoS$_4$ bilayer is still an AM, because the absence of the $[C_2\|\mathcal{P}]$ and $[C_2\|\mathcal{M}_z]$ symmetries enforces the spin splitting and the presence of the $[C_2\|\mathcal{M}_d]$ symmetry guarantees the degeneracy between the X and Y valleys with opposite spins. By contrast, the non-diagonal-sliding $Fe_2MoS_4$ bilayer breaks the $[C_2\|\mathcal{M}_d]$ symmetry that ensures the X and Y valleys belong to the same wave vector star, thereby giving rise to the valley splitting $E_\sigma(\mathbf{k}_X) \neq E_{-\sigma}(\mathbf{k}_Y)$. Despite exhibiting nonrelativistic spin splitting and zero net magnetization, the presence of spontaneous valley splitting indicates that the non-diagonal-sliding $Fe_2MoS_4$ bilayer does not satisfy the strict symmetry definition of an AM [Eq. (1)] and should instead be regarded as a fully compensated ferrimagnetic state [66]. In addition to the coexistence of spin splitting and valley splitting in the $Fe_2MoS_4$ bilayer ($\hat{O}=[E\|\mathcal{M}_z|\mathbf{t}_\parallel]$) that lacks the $[C_2\|\mathcal{P}]$, $[C_2\|\mathcal{M}_z]$ and $[C_2\|\mathcal{M}_d]$ symmetries, the coexistence of spin degeneracy and valley splitting can be realized in the $Fe_2MoS_4$ bilayer constructed via $\hat{O}=[C_2\|\mathcal{M}_z|\mathbf{t}_\parallel]$, which preserves the $[C_2\|\mathcal{P}]$ symmetry but lacks the $[C_2\|\mathcal{M}_d]$ symmetry [see Sec. VIII of SM].

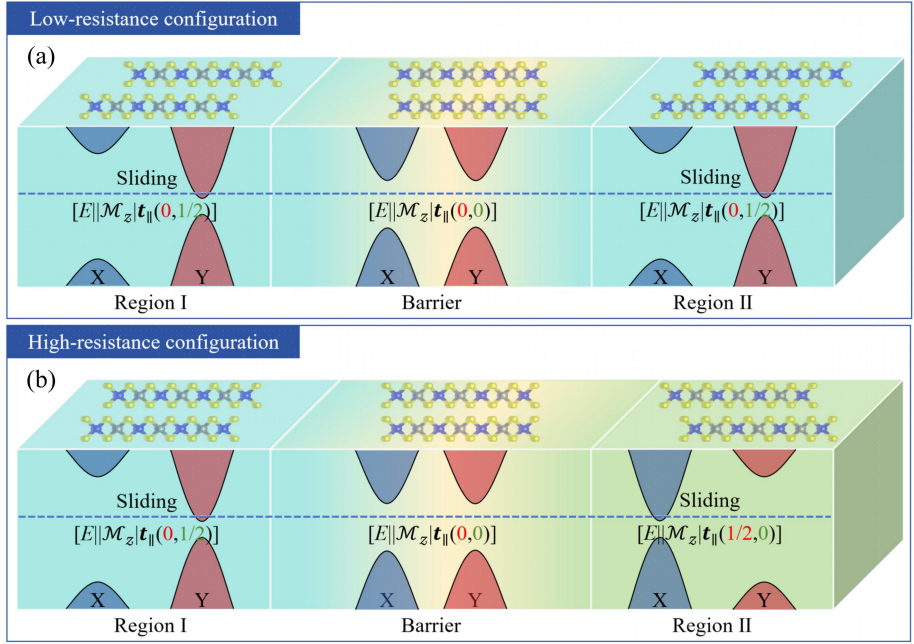

FIG. 6. Schematic of TMR device based on $Fe_2MoS_4$ bilayer ($\hat{O}=[E\|\mathcal{M}_z|\mathbf{t}_\parallel(t_a,t_b)]$). (a) The low-resistance and (b) high-resistance configurations of TMR device. The blue dashed line denotes the Fermi level.

The layer-mediated valley DOF tuning in AM bilayer provides an alternative route toward enhanced tunneling magnetoresistance (TMR) device, without relying on the external magnetic-field control [67]. As illustrated in Fig. 6, two reversed valley splitting configurations emerge from interlayer sliding of the $Fe_2MoS_4$ bilayer along $\mathbf{a}$ ($t_a = 1/2$) and $\mathbf{b}$ ($t_b = 1/2$) directions, serving as bistable states for realizing a tunable TMR response. With a suitable Fermi level, the middle region (barrier region) is still insulated while the other two regions become doped. In the low-resistance configuration, the match of the spin channels and valleys of the CBM between the region I and region II leads to a low-conductance contrast, as shown in Fig. 6(a). In contrast, beyond the conventional TMR that relies solely on spin channels alignment, the high-resistance configuration additionally exhibits a mismatch of the valley index in momentum space [see Fig. 6(b)]. Owing to the large momentum-space separation between the inequivalent X and Y valleys, electrons tunneling from region I to region II experience a substantial momentum mismatch, which strongly suppresses the tunneling probability and results in an enhanced TMR device. The low- and high-resistance configuration can be interchanged by switch spin and valley index via proper interlayer sliding.

## IV. CONCLUSIONS

In summary, symmetry analysis combined with first-principles calculations establishes a general layer-mediated strategy for tuning spin and valley DOFs in stacked tetragonal AM bilayers. In AM bilayers, the $[C_2\|\mathcal{P}]$ and $[C_2\|\mathcal{M}_z]$ symmetries inherently enforce spin degeneracy. Once these symmetries are broken, either by magnetic reconfiguration or by an external electric field, spin splitting naturally emerges. Using $V_2S_2O$ bilayers as example, we demonstrate that an external electric field couples directly to layer DOF, enabling efficient electric-field control of spin splitting through a layer-mediated mechanism. In contrast, valley physics can be tuned by interlayer sliding. Using $Fe_2MoS_4$ bilayers as example, we show that appropriate interlayer sliding breaks the $[C_2\|\mathcal{M}_d]$ symmetry and provides a new pathway for achieving tunable valley splitting, independent of conventional uniaxial strain. Enabled by the tunable spin and valley splitting, we further propose a high-contrast TMR device, in which spin-valley co-indexing leads to a pronounced suppression of tunneling channels in the high-resistance state. Our work establishes a symmetry-based and layer-mediated framework for independently engineering spin and valley DOFs in AM bilayers, highlighting promising opportunities for layer-based integration of spintronics and valleytronics.

## ACKNOWLEDGEMENTS

This work is supported by the National Natural Science Foundation of China (Grant Nos. 12274028 and 12234003) and the National Key R&D Program of China (2022YFA1402603).

## DATA AVAILABILITY

The data are available from the authors upon reasonable request.